\documentclass[english]{article}
\usepackage[T1]{fontenc}
\usepackage[latin9]{inputenc}
\usepackage{amsmath}
\usepackage{amsthm}
\usepackage[all]{xy}

\makeatletter
\newcommand{\lyxaddress}[1]{
	\par {\raggedright #1
	\vspace{1.4em}
	\noindent\par}
}
\theoremstyle{plain}
\newtheorem{thm}{\protect\theoremname}[section]

\@ifundefined{date}{}{\date{}}
\usepackage{tikz}
\newcommand*{\encircled}[1]{\relax\ifmmode\mathpalette\@encircled@math{#1}\else\@encircled{#1}\fi}
\newcommand*{\@encircled@math}[2]{\@encircled{$\m@th#1#2$}}
\newcommand*{\@encircled}[1]{%
  \tikz[baseline,anchor=base]{\node[draw,circle,outer sep=0pt,inner sep=.2ex] {#1};}}
\makeatother

\usepackage[titletoc,title]{appendix}
\usepackage{titlesec}
\usepackage{chngcntr}

\newcommand{\Proj}{\mathsf{Proj}}
\usepackage{cite}

\makeatother

\usepackage{babel}
\providecommand{\theoremname}{Theorem}

\begin{document}
\title{Hidden variables, free choice, context-independence, and all that}
\author{Ehtibar N. Dzhafarov}
\maketitle

\lyxaddress{\begin{center}
\textsuperscript{}Purdue University, USA, ehtibar@purdue.edu
\par\end{center}}
\begin{abstract}
This paper provides a systematic account of the hidden variable models
(HVMs) formulated to describe systems of random variables with mutually
exclusive contexts. Any such system can be described either by a model
with free choice but generally context-dependent mapping of the hidden
variables into observable ones, or by a model with context-independent
mapping but generally compromised free choice. These two types of
HVMs are equivalent, one can always be translated into another. They
are also unfalsifiable, applicable to all possible systems. These
facts, the equivalence and unfalsifiability, imply that freedom of
choice and context-independent mapping are no assumptions at all,
and they tell us nothing about freedom of choice or physical influences
exerted by contexts as these notions would be understood in science
and philosophy. The conjunction of these two notions, however, defines
a falsifiable HVM that describes noncontextuality when applied to
systems with no disturbance or to consistifications of arbitrary systems.
This HVM is most adequately captured by the term ``context-irrelevance,''
meaning that no distribution in the model changes with context.

\end{abstract}

\section{Introduction}

Hidden variable models (HVMs) are arguably the main reason why contextuality
and its nonlocality version have acquired prominence in the foundations
of quantum mechanics (QM). Ever since it was accepted that results
of a measurement, such as that of a spin, are almost always random
variables (with the exception of repeated sharp measurements), physicists
have been interested in the possibility of ``explaining'' such random
variables as deterministic functions of some underlying sources of
variability, even if as yet unknown to us, ``hidden.'' This possibility
is often presented as a belief famously held by Albert Einstein, and
then famously ruled out by Bell's and Kochen-Specker's theorems juxtaposed
with QM predictions \cite{Bell1966,KochenSpecker1967}.

However, even before any detailed analysis, there is a good reason
to doubt that HVMs can play an explanatory role. The reason is that
the existence of a random variable of which several jointly distributed
random variables are deterministic functions is ensured trivially:
the properties of being jointly distributed and being functions of
a single random variable are one and the same property. Conversely,
variables that are not jointly distributed, as they are predicated
on mutually exclusive conditions, cannot be functions of a single
random variable. This means that one must have as many hidden variables
as there are mutually exclusive contexts, even if they all have the
same distribution. This is not to say that HVMs cannot be meaningfully
constructed and interpreted. This only means that one should be careful
not to attach deep physical or otherwise substantive connotations
to purely mathematical and universally satisfiable representations.
This is a point elaborated throughout the paper.

In this paper I will synthesize some of my recent published work to
provide a comprehensive and rigorous account of HVMs. The most restrictive
HVM, one introduced by Bell and describing noncontextual systems with
no disturbance, is known not to hold for many systems of random variables.
When this happens, the constraints imposed on an HVM have to be relaxed,
and this can be done in two ways: either by allowing for a dependence
of the measurement outcome distributions on contexts or by allowing
for an interdependence between the hidden variables and the choices
of settings for the measurements. In Ref. \cite{Dzh2022} I proved
the equivalence of these two options. In this paper I present an improved
and more rigorous proof. I will argue that such assumptions as freedom
of choice and context-independent mapping (of hidden variables into
observable ones) are merely metaphorical depictions of some basic
representations of jointly distributed random variables. Next, I discuss
the problem of separating disturbance (or signaling) from contextuality
in the situations in which Bell's HVM does not hold. While this is
the central issue for the Contextuality-by-Default (CbD) theory \cite{Elephants,DzhCerKuj2017,KD2021,DzhKuj2023Nonmeasurements},
the difference between disturbance and contextuality is not apparent
in the formulations of the HVMs. However, one can effectively separate
disturbance from contextuality by using the consistified systems introduced
in Refs. \cite{DK2023,chapter}. Any system of random variables can
be reformulated as an equivalent, in a well-defined sense, system
that has no disturbance (is consistently connected, in the CbD terminology).
The equivalence of the HVMs with context-dependent mapping and the
HVMs with violations of free choice holds for these consistified systems
too, but now any such HVM indicates pure contextuality. At the conclusion
of the paper I will discuss two assumptions that one could suspect
to be required for the development presented, and show that, once
again, they are not assumptions at all, because they are trivially
satisfied in the language of random variables.

\section{Conceptual and terminological set-up}

A \emph{system of random variables} is a double-indexed set

\begin{equation}
\mathcal{R}=\left\{ R_{q}^{c}:c\in C,q\in Q,q\prec c\right\} ,\label{eq:system}
\end{equation}
where $Q$ is a set of \emph{contents}, $C$ is a set of \emph{contexts},
and $q\prec c$ means that content $q$ is measured in context $c$.
A content $q$ in $R_{q}^{c}$ can be viewed as a question that the
random variable $R_{q}^{c}$ answers (e.g., ``is the spin along axis
$q$ up?'', answered ``yes/no'') or as a choice of measurements
(``spin along axis $q$'') whose outcomes (``up/down'') are represented
by $R_{q}^{c}$. The context $c$ in $R_{q}^{c}$ indicates conditions
under which $R_{q}^{c}$ is recorded, such as the set of all other
measurements made together with $R_{q}^{c}$ and the spatial and temporal
relations among them. The matrix below provides an example of a system
of random variables:
\begin{equation}
\begin{array}{|c|c|c|c||c|}
\hline R_{1}^{1} & R_{2}^{1} & R_{3}^{1} &  & c=1\\
\hline  &  & R_{3}^{2} & R_{4}^{2} & c=2\\
\hline R_{1}^{3} & R_{2}^{3} & R_{3}^{3} & R_{4}^{3} & c=3\\
\hline\hline q=1 & q=2 & q=3 & q=4 & \textnormal{system }\mathcal{R}_{0}
\\\hline \end{array}\,.\label{eq:example}
\end{equation}
The subsystem of all random variables within a given context $c$
is called a \emph{bunch} (of random variables),
\begin{equation}
R^{c}=\left\{ R_{q}^{c}:q\in Q^{c}\right\} ,
\end{equation}
where 
\begin{equation}
Q^{c}=\left\{ q\in Q:q\prec c\right\} .
\end{equation}
For instance, the bunch $R^{2}$ in the system $\mathcal{R}_{0}$
is $\left\{ R_{3}^{2},R_{4}^{2}\right\} $. Any bunch $R^{c}$ is
a random variable, which means that all components $R_{q}^{c}$ of
$R^{c}$ are jointly distributed (are measurable functions on the
same probability space). However, no two random variables from different
bunches have a joint distribution, they are \emph{stochastically unrelated}
(are measurable functions on distinct probability spaces). Indeed,
consider what a joint distribution of $R_{q}^{c}$ and $R_{q'}^{c'}$
with $c\not=c'$ could look like (X and Y being any measurable sets):
\begin{equation}
\begin{array}{|c|c|c|c|}
\hline  & R_{q'}^{c'}\in Y & R_{q'}^{c'}\not\in Y & \\
\hline R_{q}^{c}\in X & ? & ? & \\
\hline R_{q}^{c}\not\in X & ? & ? & \\
\hline  &  &  & 1
\\\hline \end{array}\,.
\end{equation}
The question marks cannot be all replaced with zeros because they
must sum to one. At the same time, any nonzero joint probability would
indicate that $R_{q}^{c}$ and $R_{q'}^{c'}$ co-occur, which would
contradict the fact that $c$ and $c'$ are mutually exclusive contexts.

\section{Hidden variable models}

\subsection{\label{subsec:(Excessively)-general-HVM}(Excessively) general HVM}

Let us begin with the most general possible HVM, denoted $\mathsf{HVM_{Gen}}$:

\begin{equation}
R^{c}=\alpha\left(Q^{c},\Lambda^{c}\left(c\right),c\right).\label{eq:HVMGen}
\end{equation}
The function $\alpha$ returns as its value an indexed set, and the
dependence of $\alpha$ on $Q^{c}$ should be understood as its indexing,
matching the indexing of $R^{c}$. Thus, for system $\mathcal{R}_{0}$
in (\ref{eq:example}), $Q^{2}=\left\{ 3,4\right\} $, and the $\mathsf{HVM_{Gen}}$
representation for $R^{2}=\left(R_{3}^{2},R_{4}^{2}\right)$ is
\begin{equation}
\begin{array}{r}
\alpha\left(Q^{2},\Lambda^{2}\left(2\right),2\right)=\left(\Proj_{q=3}\alpha\left(Q^{2},\Lambda^{2}\left(2\right),2\right),\Proj_{q=4}\alpha\left(Q^{2},\Lambda^{2}\left(2\right),2\right)\right)\\
=\left(\alpha\left(3,\Lambda^{2}\left(2\right),2\right),\alpha\left(4,\Lambda^{2}\left(2\right),2\right)\right),
\end{array}
\end{equation}
where $\Proj_{q}V$ stands for the $q$-indexed component of the indexed
set $V$.\footnote{The notation $\Lambda^{c}\left(c\right)$ may appear excessive, but
it is not. The superscript $c$ merely indicates that the random variables
in different contexts are different and stochastically unrelated (this
applies to both $\Lambda^{c}$ and $R^{c}$). The superscript $c$
therefore is universal and ineliminable. The dependence of $\Lambda^{c}$
on $c$ as an argument means that the \emph{distribution} of the hidden
variable may be different in different contexts. This may or may not
be the case in other HVMs.}

One can present this model graphically as

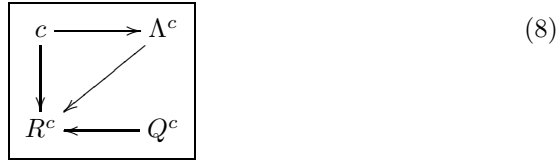
\begin{equation}
\boxed{\xymatrix@C=1cm{c\ar[r]\ar[d] & \Lambda^{c}\ar[dl]\\
R^{c} & Q^{c}\ar[l]
}
}
\end{equation}
The arrows $a\rightarrow b$ in this and subsequent diagrams (where
$b$ is a random variable and $a$ is a random variable or a parameter)
should be read as ``different values of $a$ may result in different
distributions of $b$.''

$\mathsf{HVM_{Gen}}$ is not a falsifiable model, it can be applied
to any system of random variables. This can be demonstrated by simply
putting $\Lambda^{c}\left(c\right)=R^{c}$, with the stand-alone $c$
in $\alpha$ becoming a dummy argument, and $Q^{c}$ extracted from
$\Lambda^{c}\left(c\right)$ as its indexing set.

\subsection{\label{subsec::-Context-independence}Context-independent mapping
without free choice}

The argument just presented shows that the direct dependence of the
distribution of $R^{c}$ on $c$ can be eliminated:
\begin{equation}
R^{c}=\beta\left(Q^{c},\Lambda^{c}\left(c\right)\right),\label{eq:HVMCntInd}
\end{equation}
or graphically,
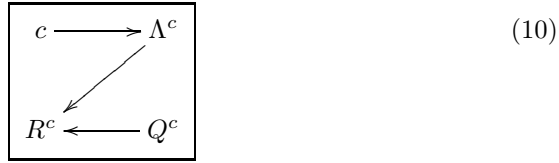
\begin{equation}
\boxed{\xymatrix@C=1cm{c\ar[r] & \Lambda^{c}\ar[dl]\\
R^{c} & Q^{c}\ar[l]
}
}
\end{equation}
Although not obvious at first glance, this HVM would traditionally
be interpreted as a model with a context-independent mapping of $\Lambda^{c}\left(c\right)$
into $R^{c}$ (no arrow from $c$ to $R^{c}$) but with generally
compromised freedom of choice (the distribution of $\Lambda^{c}$
may depend on $c$).

I will denote this model $\mathsf{HVM_{+CIM}^{-FC}}$, using the self-evident
abbreviations. We have established that $\mathsf{HVM_{Gen}}$ can
always be reduced to $\mathsf{HVM_{+CIM}^{-FC}}$. Using again as
an example system $\mathcal{R}_{0}$ in (\ref{eq:example}), the $\mathsf{HVM_{+CIM}^{-FC}}$
representation for $R^{2}=\left(R_{3}^{2},R_{4}^{2}\right)$ is
\begin{equation}
\beta\left(Q^{2},\Lambda^{2}\left(2\right)\right)=\left(\beta\left(3,\Lambda^{2}\left(2\right)\right),\beta\left(4,\Lambda^{2}\left(2\right)\right)\right).
\end{equation}

Freedom of choice in the QM literature is usually discussed in terms
of the relationship between one's choice of $c$ and the hidden variable
$\Lambda^{c}\left(c\right)$. This means that $c$ is treated as a
random variable (which is a dubious viewpoint, see Ref. \cite{Dzh2022}),
and freedom of choice means that $c$ and $\Lambda^{c}$ are stochastically
independent. In my representation of HVMs, $c$ is always a deterministic
parameter, which, with respect to the traditional view, simply means
that all random variables in the model are conditioned on fixed values
of $c.$ Any restriction of freedom of choice in the traditional sense
then translates into a dependence of the distribution of $\Lambda^{c}$
on $c$. As a special case, this also applies to the possibility that
$c$ is a function of the hidden variable, $c=f\left(\Lambda^{c}\right)$,
which may possibly be interpreted as a depiction of superdeterminism:.
in an $\mathsf{HVM_{+CIM}^{-FC}}$, one simply replaces this function
with $\Lambda^{c}\left(c\right)$, defined by $f\left(\Lambda^{c}\left(c\right)\right)=c$.

\subsection{\label{subsec::-Free-choice}Free choice without context-independent
mapping}

It is further possible to transform $\mathsf{HVM_{+CIM}^{-FC}}$ into
a model that is, in a sense, its reverse. Given (\ref{eq:HVMCntInd}),
one can form an arbitrary coupling of $\Lambda^{c}$ for all contexts
$c$,\footnote{This coupling is a random variable whose components are indexed by
$c$ and distributed as the corresponding $\Lambda^{c}$.}
\begin{equation}
\Gamma:=\left\{ \Lambda^{c}\left(c\right):c\in C\right\} ,
\end{equation}
and then create, for every $c\in C$, a distributional copy $\Gamma^{c}$
of $\Gamma$, so that these copies are pairwise stochastically unrelated.
Then
\begin{equation}
\Lambda^{c}\left(c\right)=\Proj_{c}\Gamma^{c},
\end{equation}
and
\begin{equation}
R^{c}=\beta\left(Q^{c},\Proj_{c}\Gamma^{c}\right)=\gamma\left(Q^{c},\Gamma^{c},c\right),\label{eq:HVMFCCD}
\end{equation}
where the variables $\Gamma^{c}$ (as indicated by the lack of $c$
as their argument) have one and the same distribution for all $c\in C$.
Note that we cannot eliminate the index $c$ in $\Gamma^{c}$, because
$R^{c}=\gamma\left(Q^{c},\Gamma,c\right)$ would make all $R^{c}$
jointly distributed.

The traditional interpretation of the HVM described by (\ref{eq:HVMFCCD})
would be that the freedom of choice is not compromised here, but context-independence
is generally violated. Using our graphical representation,
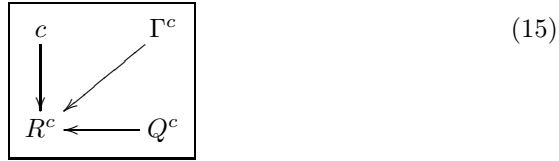
\begin{equation}
\boxed{\xymatrix@C=1cm{c\ar[d] & \Gamma^{c}\ar[dl]\\
R^{c} & Q^{c}\ar[l]
}
}
\end{equation}

I will denote this model $\mathsf{HVM_{-CIM}^{+FC}}$. We have established
that $\mathsf{HVM_{+CIM}^{-FC}}$ implies (can be translated into)
$\mathsf{HVM_{-CIM}^{+FC}}$. Using our example of system $\mathcal{R}_{0}$
in (\ref{eq:example}), the $\mathsf{HVM_{-CIM}^{+FC}}$ representation
for $R^{2}=\left(R_{3}^{2},R_{4}^{2}\right)$ is
\begin{equation}
\gamma\left(Q^{2},\Gamma^{2},2\right)=\left(\gamma\left(3,\Gamma^{2},2\right),\gamma\left(4,\Gamma^{2},2\right)\right).
\end{equation}

\subsection{Free choice with context-independent mapping}

Both $\mathsf{HVM_{+CIM}^{-FC}}$, and $\mathsf{HVM_{-CIM}^{+FC}}$
can be viewed as deviations from their special case

\begin{equation}
R^{c}=\delta\left(Q^{c},\Gamma^{c}\right),\label{eq:HVMBell}
\end{equation}
or, graphically, 
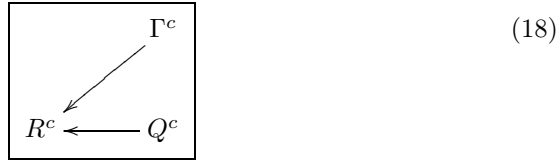
\begin{equation}
\boxed{\xymatrix@C=1cm{ & \Gamma^{c}\ar[dl]\\
R^{c} & Q^{c}\ar[l]
}
}
\end{equation}
where the random variables $\Gamma^{c}$ for all $c\in C$ are identically
distributed and pairwise stochastically unrelated. This model can
be denoted $\mathsf{HVM_{+CIM}^{+FC}}$, as it satisfies both freedom
of choice and context-independence in the mapping of $\Gamma^{c}$
into $R^{c}$. In our example of system $\mathcal{R}_{0}$ in (\ref{eq:example}),
the $\mathsf{\mathsf{HVM_{+CIM}^{+FC}}}$ representation for $R^{2}=\left(R_{3}^{2},R_{4}^{2}\right)$
is
\begin{equation}
\delta\left(Q^{2},\Gamma^{2}\right)=\left(\delta\left(3,\Gamma^{2}\right),\delta\left(4,\Gamma^{2}\right)\right).
\end{equation}

Unlike the previous two HVMs, this one is a true model, as it is falsifiable.
The latter is demonstrated, for instance, by relating predictions
of QM to the Bell-type \cite{Bell1966} and Kochen-Specker-type theorems
(in addition to the original Refs. \cite{Bell1966,KochenSpecker1967}
see, e.g., Refs. \cite{CHSH1969,Cabello2008,Fine1982}). The Bell-type
theorems establish necessary and sufficient conditions for a system
of random variables to be described by $\mathsf{HVM_{+CIM}^{+FC}}$,
which can then be shown to fail for some QM systems. In the Kochen-Specker-type
theorems one constructs systems of random variables in accordance
with QM, and then demonstrate that they cannot be described by $\mathsf{HVM_{+CIM}^{+FC}}$.

\subsection{Equivalence theorem and its consequences}

Combining the implications in Subsections \ref{subsec::-Context-independence}
and \ref{subsec::-Free-choice}, 
\begin{equation}
\mathsf{HVM_{Gen}}\Rightarrow\mathsf{HVM_{+CIM}^{-FC}}\Rightarrow\mathsf{HVM_{-CIM}^{+FC}},
\end{equation}
and observing that $\mathsf{HVM_{-CIM}^{+FC}}$ is a special case
of $\mathsf{HVM_{Gen}}$, we obtain the following statement.
\begin{thm}
\textup{The models $\mathsf{HVM_{Gen}}$, $\mathsf{HVM_{+CIM}^{-FC}}$,
and $\mathsf{HVM_{-CIM}^{+FC}}$ are pairwise equivalent:
\begin{equation}
\vcenter{\xymatrix@C=1cm{ & \mathsf{HVM_{Gen}}\ar@{<=>}[dl]\\
\mathsf{\mathsf{HVM_{+CIM}^{-FC}}\ar@{<=>}[rr]} &  & \mathsf{HVM_{-CIM}^{+FC}}\ar@{<=>}[ul]
}
}
\end{equation}
}
\end{thm}
Let us consider two consequences of this theorem. One of them is that
when $\mathsf{HVM_{+CIM}^{+FC}}$ is not applicable to a system, one
can arbitrarily choose between describing the system in the language
of $\mathsf{HVM_{+CIM}^{-FC}}$ or in the language of $\mathsf{HVM_{-CIM}^{+FC}}$.
In particular, one can always use one and the same measure for the
degree of deviation of these two HVMs from $\mathsf{HVM_{+CIM}^{+FC}}$:
\begin{equation}
\xymatrix@C=1cm{\mathsf{HVM_{-CIM}^{+FC}}\ar@{<-->}[rr]^{deviation} &  & \mathsf{HVM_{+CIM}^{+FC}}\ar@{<-->}[rr]^{deviation} &  & \mathsf{HVM_{+CIM}^{-FC}}}
\end{equation}
A special case of this corollary, for a particular system of random
variables, is presented in Ref. \cite{Blasiaketal.2021}.

The second consequence of the theorem is that $\mathsf{HVM_{+CIM}^{-FC}}$
and $\mathsf{HVM_{-CIM}^{+FC}}$ are both unfalsifiable, either of
them can describe any system of random variables. This follows from
the demonstration, at the end of Section \ref{subsec:(Excessively)-general-HVM},
that $\mathsf{HVM_{Gen}}$ is unfalsifiable, in fact, even in the
form of $\mathsf{HVM_{+CIM}^{-FC}}$. This, in combination with the
inter-translatability of $\mathsf{HVM_{+CIM}^{-FC}}$ and $\mathsf{HVM_{-CIM}^{+FC}}$,
should make one skeptical about interpreting the dependence of the
distribution of $\Lambda^{c}$ on $c$ in terms of ``freedom of choice,''
in any substantive meaning of these words, and interpreting an arrow
from $c$ to $R^{c}$ as a physical influence exerted by the context.
Their complete equivalence and empirical emptiness (universal applicability)
suggest the view that $\mathsf{HVM_{+CIM}^{-FC}}$ and $\mathsf{HVM_{-CIM}^{+FC}}$
are purely mathematical descriptions of the joint distributions within
bunches of random variables and of the differences between them.

This view does not change if one constrains or even completely specifies
all distributions and functions in the formulation of $\mathsf{HVM_{+CIM}^{-FC}}$
or $\mathsf{HVM_{-CIM}^{+FC}}$, making them thereby predictive and
falsifiable. The inter-translatability of the two types of models
holds irrespective of their falsifiability. Moreover, a completely
specified HVM can always be thought of as a corresponding unconstrained
HVM after it has been applied to the system predicted by the completely
specified HVM. Clearly, the ontological interpretation of a model
(say, $\mathsf{HVM_{+CIM}^{-FC}}$) does not depend on whether it
has been applied to a particular system of random variables, because
this does not change the facts that (A) it could have been applied
to any other system, and (B) it can be translated into an HVM of a
completely different nature (in this case, $\mathsf{HVM_{-CIM}^{+FC}}$).

This is not to say that the notions of freedom of choice and context-(in)dependent
mapping may not be assigned substantive meanings and be propitiously
used in physical or other scientific theories. One should, however,
distinguish HVMs per se from scientific theories that predict specific
systems of random variables and therefore HVM representations thereof.
My only point here is that these substantive meanings belong to the
parts of theories extraneous to the HVMs the theories lead to. In
other words, these meanings cannot be derived from the HVMs themselves,
from the fact that a system can be described by $\mathsf{HVM_{+CIM}^{-FC}}$
or $\mathsf{HVM_{-CIM}^{+FC}}$ (or even $\mathsf{HVM_{Gen}}$, combining
the two) --- because any system can, and by any of them. The language
of HVMs as understood in this paper (and in most discussions of the
HVMs in the foundations of physics, beginning with Bell's work), is
simply too crude to capture certain substantive notions and distinctions.
(We will see below that it is sometimes too crude even to depict the
difference between much more clear-cut notions of contextuality and
signaling.) A simple analogy may help to understand this. Any real-valued
random variable $R$ can be generated by applying an appropriate transformation
$f$ to a variable $U$ uniformly distributed between 0 and 1. As
one observes values of $R$, it is possible that there is a computer
program that de facto computes them by first generating values of
$U$ and then applying to them the function $f$. If this is known
from some extraneous source of knowledge, then we have a valid naturalistic
interpretation of the model $R=f\left(U\right)$, which then acquires
a privileged status over other representations of $R$ (such as $R=g\left(E\right)$,
for an exponentially distributed $E$). However, such an interpretation
cannot be derived from the fact that $R$ is representable as $f\left(U\right)$
--- because this representation is mathematically guaranteed, and
moreover, is not unique (referring, e.g., to the same $R=g\left(E\right)$).

The terms ``freedom of choice'' and ``context-(in)dependent mapping''
may still be conveniently used as labels for HVM components, provided
one does not impute to them their colloquial, physical, or philosophical
connotations. Moreover, the conjunction of these two notions does
have a substantive meaning, because $\mathsf{HVM_{+CIM}^{+FC}}$ is
a falsifiable model which de facto does not apply to some QM systems
of random variables. In Ref. \cite{Dzh2022} I argued that the notions
in question should only be used in conjunction: ``one cannot accept
local causality without free choice, because denying free choice is
equivalent to denying local causality'' (local causality being the
specific form of context-independent mapping used by Bell in the discussion
published in Ref. \cite{BellvsSHC1985}). While the present paper
only strengthens this assertion, I would like to add here that one
can very well decide to abandon the terms ``freedom of choice''
and ``context-(in)dependent mapping'' altogether, and use instead
a simpler way to characterize $\mathsf{HVM_{+CIM}^{+FC}}$. Namely,
this is the model in which context $c$ is irrelevant for determining
any distributions involved (which includes the distribution of the
hidden variable $\Lambda^{c}$ and the distribution of the observable
bunch $R^{c}$). Therefore, $\mathsf{HVM_{+CIM}^{+FC}}$ can be referred
to as the the model satisfying the assumption of \emph{context-irrelevance}.

\section{Contextuality in consistently connected systems}

We have managed so far to discuss HVMs without involving the notion
of \emph{(non)contextuality}. It is now time to involve it. The traditional
definition of this notion simply coincides with that of $\mathsf{HVM_{+CIM}^{+FC}}$:
a system of random variables is noncontextual (or, for distributed
systems, local) if it is described by this HVM; and a system that
cannot be so described is contextual. One consequence of this definition
is that a noncontextual system must be \emph{consistently connected}.
The latter is a CbD term for what is usually called in QM \emph{non-disturbance}
or \emph{non-signaling}: in a consistently connected system, any two
random variables sharing a content, $R_{q}^{c}$ and $R_{q}^{c'}$,
have the same distribution. \emph{Inconsistent connectedness} (disturbance,
signaling) therefore makes a system contextual. This definition makes
the class of contextual systems too large and heterogeneous, and CbD
offers a more analytic approach, presented in the next section. For
now, however, let us confine consideration to consistently connected
systems.\footnote{In fact, $\mathsf{HVM_{+CIM}^{+FC}}$ predicts a more restricted form
of consistent connectedness, termed in CbD \textit{strong consistent
connectedness}: for any contexts $c,c'$ and any set of contents $I\subseteq Q^{c}\cap Q^{c'}$,
we should have identically distributed $\left\{ R_{q}^{c}:q\in I\right\} $
and $\left\{ R_{q}^{c'}:q\in I\right\} $. A system that is consistently
connected but not strongly so is always contextual in CbD.}

The main consequence of $\mathcal{R}$ being described by $\mathsf{HVM_{+CIM}^{+FC}}$
is as follows. With reference to (\ref{eq:HVMBell}), construct the
random variable $S$ defined by 
\begin{equation}
S=\delta\left(Q,\Gamma\right),\label{eq:reduced couling}
\end{equation}
where $\Gamma$ has the same distribution as $\Gamma^{c}$ in (\ref{eq:HVMBell}).
The variable $S$ is called a \emph{reduced coupling} of the system
$\mathcal{R}$ \cite{DK2016}. Its (jointly distributed) elements
are indexed by the elements of $Q$, and for any $c\in C$, we have
\begin{equation}
R^{c}\overset{d}{=}\Proj_{Q^{c}}S,\label{eq:subject to}
\end{equation}
where $\overset{d}{=}$ indicates equality of distributions. Thus,
for our system $\mathcal{R}_{0}$ in (\ref{eq:example}), the reduced
coupling has the form $S=\left\{ S_{1},S_{2},S_{3},S_{4}\right\} $,
and the condition (\ref{eq:subject to}) means that in the matrix

\begin{equation}
\begin{array}{|c|c|c|c||c|}
\hline S_{1} & S_{2} & S_{3} &  & c=1\\
\hline  &  & S_{3} & S_{4} & c=2\\
\hline S_{1} & S_{2} & S_{3} & S_{4} & c=3\\
\hline\hline q=1 & q=2 & q=3 & q=4 & 
\\\hline \end{array}\label{eq:coupling for exmaple}
\end{equation}
the rows are distributed as the corresponding rows in (\ref{eq:example}).

It is clear that the implication $\mathsf{HVM_{+CIM}^{+FC}}\Rightarrow S$
can be reversed, whence we have the following criterion: system $\mathcal{R}$
is described by $\mathsf{HVM_{+CIM}^{+FC}}$ if and only if it has
a reduced coupling (\ref{eq:reduced couling}) subject to (\ref{eq:subject to}).
For some simple systems this has been semi-formally derived as the
``joint distribution criterion'' by Fine \cite{Fine1982}, based
on the idea of Suppes and Zanotti \cite{SuppesZanotti1981}. Note
that the use of the language of random variables makes this criterion
obtain essentially automatically.

For these and other simple systems (notably for the important class
of so-called \emph{cyclic systems} \cite{cyclic2013}) other criteria
have been derived, primarily in the form of inequalities involving
expected values of the products of the random variables within different
bunches. These additional criteria should be viewed as mere shortcuts,
because in all cases when they are available and in many cases when
they are not, the existence or non-existence of a reduced coupling
(\ref{eq:HVMBell})-(\ref{eq:subject to}) can be established directly,
by means of linear programing.

This is a good place to note that some authors, having correctly observed
that Bell-type inequalities require a system of jointly distributed
variables, as in (\ref{eq:coupling for exmaple}), and having also
correctly observed that in a system of observable probabilities different
bunches are not jointly distributed, have then erroneously concluded
that the Bell-type theorems were wrong \cite{Khrennikov2009,Khrennikov2015,Kupzc2021}.
In fact, the only problem with these theorems, from the earliest ones
in the 1960s all the way to the present, is that they are usually
proved less than rigorously, with unacknowledged abuse of notation.
When viewed as theorems about reduced couplings, their proofs are
correct. The corrected proofs do not require that different bunches
be jointly distributed. They only require that a system can be described
by $\mathsf{HVM_{+CIM}^{+FC}}$, the model that does preserve stochastic
unrelatedness of different bunches.

\section{Contextuality in inconsistently connected systems}

CbD offers a generalized notion of (non)contextuality, one that applies
to all systems of random variables, including \emph{inconsistently
connected} ones (those with disturbance, or signaling).\footnote{More precisely, the current, second version of CbD, applies to arbitrary
systems of \emph{dichotomous} random variables \cite{DzhCerKuj2017,KD2021}.
However, this constraint is not relevant to the present discussion.
Moreover, the discussion below could easily be generalized to a class
of approaches that include CbD as a special case \cite{DK2023}.} Given a system $\mathcal{R}$ in (\ref{eq:system}), its (complete)
\emph{coupling} is defined as a random variable 
\begin{equation}
S=\left\{ S_{q}^{c}:c\in C,q\in Q,q\prec c\right\} 
\end{equation}
such that, for every $c\in C$,
\begin{equation}
S^{c}\overset{d}{=}R^{c},
\end{equation}
where 
\begin{equation}
S^{c}:=\left\{ S_{q}^{c}:q\in Q^{c}\right\} .
\end{equation}
Note that calling $S$ a random variable implies that, unlike in the
system $\mathcal{R}$, all components of $S$ are jointly distributed.
A system $\mathcal{R}$ is noncontextual if it has a coupling $S$
in which, for every content $q\in Q$ and any two contexts $c,c'$
such that $q\prec c$ and $q\prec c'$, the probability 
\begin{equation}
p\left[S_{q}^{c}=S_{q}^{c'}\right]
\end{equation}
is maximal possible. The maximum is computed for fixed distributions
of $S_{q}^{c},S_{q}^{c'}$ (which coincide with the distributions
of $R_{q}^{c},R_{q}^{c'}$, respectively). If such a coupling does
not exist, $\mathcal{R}$ is contextual. If $\mathcal{R}$ is consistently
connected, then the maximal probability for any event $S_{q}^{c}=S_{q}^{c'}$
equals 1, and the definition reduces to the existence of the reduced
coupling introduced in the previous section.

To illustrate this for our example (\ref{eq:example}), a coupling
for $\mathcal{R}_{0}$ is a random variable 
\begin{equation}
S=\begin{array}{|c|c|c|c||c|}
\hline S_{1}^{1} & S_{2}^{1} & S_{3}^{1} &  & c=1\\
\hline  &  & S_{3}^{2} & S_{4}^{2} & c=2\\
\hline S_{1}^{3} & S_{2}^{3} & S_{3}^{3} & S_{4}^{3} & c=3\\
\hline\hline q=1 & q=2 & q=3 & q=4 & 
\\\hline \end{array}
\end{equation}
whose rows are distributed as the corresponding rows in (\ref{eq:example}).
$\mathcal{R}_{0}$ is noncontextual if and only if its coupling $S$
can be chosen so that the probabilities of the events
\begin{equation}
S_{1}^{1}=S_{1}^{3},S_{2}^{1}=S_{2}^{3},S_{3}^{1}=S_{3}^{2},S_{3}^{1}=S_{3}^{3},S_{3}^{2}=S_{3}^{3},S_{4}^{2}=S_{4}^{3}
\end{equation}
are all maximal possible. In particular, if $\mathcal{R}_{0}$ is
consistently connected, then it is noncontextual if and only if all
these probabilities in some coupling $S$ equal 1. In such a coupling,
the variables $S_{1}^{1}$ and $S_{1}^{3}$ can both be renamed into
$S_{1}$, the variables $S_{2}^{1}$ and $S_{2}^{3}$ can be renamed
into $S_{2}$, etc. We thus obtain the reduced coupling $\left\{ S_{1},S_{2},S_{3},S_{4}\right\} $
subject to (\ref{eq:coupling for exmaple}).

\section{Consistified systems}

What is an HVM representation of contextuality in the case when a
system may be inconsistently connected? Clearly, Bell's $\mathsf{HVM_{+CIM}^{+FC}}$cannot
be used, so one should choose between the two equivalent options:
$\mathsf{HVM_{-CIM}^{+FC}}$ and $\mathsf{HVM_{+CIM}^{-FC}}$. The
problem here is that these representations do not allow us to separate
inconsistent connectedness from contextuality. It may seem therefore
that unlike the traditional theory of contextuality, CbD cannot use
HVMs as a useful descriptive tool.

However, this difficulty can be easily remedied if one replaces a
system under consideration with its \emph{consistified} equivalent
\cite{DK2023,chapter}. A consistified equivalent $\mathcal{R}^{\dagger}$
of a system $\mathcal{R}$ is a consistently connected system that
depicts the same empirical or theoretical situation and is contextual
if and only if $\mathcal{R}$ is contextual. Specifically, given $\mathcal{R}$
in (\ref{eq:system}), $\mathcal{R}^{\dagger}$ is defined as 
\begin{equation}
\mathcal{R}^{\dagger}=\left\{ R_{\xi}^{\pi}:\pi\in C^{\dagger},\xi\in Q^{\dagger},\xi\prec^{\dagger}\pi\right\} ,\label{eq:consistified}
\end{equation}
where 
\begin{equation}
C^{\dagger}=\left\{ \pi:\pi=\left(\cdot,c\right),c\in C\right\} \sqcup\left\{ \pi:\pi=\left(q,\cdot\right),q\in Q\right\} ,
\end{equation}
\begin{equation}
Q^{\dagger}=\left\{ \xi:\xi=\left(q,c\right),q\in Q,c\in C,q\prec c\right\} ,
\end{equation}
\begin{equation}
\xi\prec^{\dagger}\pi\Longleftrightarrow\xi=\left(q,c\right)\in Q^{\dagger}\:\&\:\left[\pi=\left(\cdot,c\right)\textnormal{ or }\pi=\left(q,\cdot\right)\right].
\end{equation}
For any context $\pi=\left(\cdot,c\right)$, the bunch in this context
is defined as 
\begin{equation}
R^{\dagger\pi}=R^{\dagger\left(\cdot,c\right)}\overset{d}{=}R^{c}.
\end{equation}
To define the bunch for a context $\pi=\left(q,\cdot\right)$, we
need an auxiliary notion. For a given $q\in Q$, define a random variable
\begin{equation}
T_{q}=\left\{ T_{q}^{c}:c\in C,q\prec c\right\} ,
\end{equation}
such that for any two components $T_{q}^{c},T_{q}^{c'}$ in $T_{q}$,
\begin{equation}
T_{q}^{c}\overset{d}{=}R_{q}^{c},
\end{equation}
and the probability
\begin{equation}
p\left[T_{q}^{c}=T_{q}^{c'}\right]
\end{equation}
is maximal possible. Let us assume, for simplicity, that such $T_{q}$
exists and is unique for all $q\in Q$.\footnote{Ref. \cite{DK2023} provides an outline of how the discussion should
be modified if this is not the case. If there is more than or less
than one $T_{q}$ for some of the $q\in Q$, one should consider a
class of consistified systems $\mathcal{R}^{\dagger}$, each with
one possible combination of the realizations of $T_{q}$. This class
is deemed noncontextual if and only if at least one of its elements
is noncontextual (and this happens if and only if the original system
$R$ is noncontextual in the CbD sense). In particular, if the class
of $T_{q}$ is empty for some $q$, then the class of $\mathcal{R}^{\dagger}$
is empty, and it should be deemed contextual.} Then, for any context $\pi=\left(q,\cdot\right)$, the bunch in this
context is defined as
\begin{equation}
R^{\dagger\pi}=R^{\dagger\left(q,\cdot\right)}\overset{d}{=}T_{q}\,.
\end{equation}
This completes the construction of $\mathcal{R}^{\dagger}$.

Clearly, a consistified system is (strongly) consistently connected:
for any $\xi=\left(q,c\right)$ it contains two distributional copies
of $R_{q}^{c}$, in the contexts $\left(\cdot,c\right)$ and $\left(q,\cdot\right)$.
It should also be clear, by comparing the CbD definition of (non)contextuality
with the traditional definition applied to the consistified equivalent
of a system, that the system and its equivalent are always contextual
or noncontextual together. For a more rigorous argument, see Ref.\cite{DK2023}.

For our example $\mathcal{R}_{0}$ in (\ref{eq:example}), the consistified
equivalent is (omitting commas and brackets to save space)
\begin{equation}
\begin{array}{|c|c|c|c|c|c|c|c|c||c|}
\hline R_{11}^{\cdot1} & R_{21}^{\cdot1} & R_{31}^{\cdot1} &  &  &  &  &  &  & \pi=\cdot1\\
\hline  &  &  & R_{32}^{\cdot2} & R_{42}^{\cdot2} &  &  &  &  & \cdot2\\
\hline  &  &  &  &  & R_{13}^{\cdot3} & R_{23}^{\cdot3} & R_{33}^{\cdot3} & R_{43}^{\cdot3} & \cdot3\\
\hline R_{11}^{1\cdot} &  &  &  &  & R_{13}^{1\cdot} &  &  &  & \pi=1\cdot\\
\hline  & R_{21}^{2\cdot} &  &  &  &  & R_{23}^{2\cdot} &  &  & 2\cdot\\
\hline  &  & R_{31}^{3\cdot} & R_{32}^{3\cdot} &  &  &  & R_{33}^{3\cdot} &  & 3\cdot\\
\hline  &  &  &  & R_{42}^{4\cdot} &  &  &  & R_{43}^{4\cdot} & 4\cdot\\
\hline\hline \xi=11 & 21 & 31 & 32 & 42 & 13 & 23 & 33 & 43 & \mathcal{R}_{0}^{\dagger}
\\\hline \end{array}\,,\label{eq: exmaple consistified}
\end{equation}
where the bunches in the first three rows are distributional copies
of the corresponding rows in $\mathcal{R}_{0}$, the distributions
of the two variables in each column are identical, and in each of
the last four rows the probability of the pairwise equality of its
elements is maximal possible.

\section{Equivalence theorem for consistified systems}

The main reason why the notion of a consistified systems is useful
is the fact that the inconsistent connectedness of a system $\mathcal{R}$
is eliminated in $\mathcal{R}^{\dagger}$ (more precisely, translated
into the structure of its $\left(q,\cdot\right)$-bunches) while its
contextuality status is preserved. One can ascertain therefore whether
$\mathcal{R}^{\dagger}$ is describable by $\mathsf{HVM_{+CIM}^{+FC}}$
as one can with any other consistently connected system. If it is
not, then $\mathcal{R}^{\dagger}$ should be described by either of
$\mathsf{HVM_{-CIM}^{+FC}}$ and $\mathsf{HVM_{+CIM}^{-FC}}$, and
this time there can be no confusion as to whether they depict inconsistent
connectedness or contextuality --- it is definitely the latter. However,
the applicability of and deviations from $\mathsf{HVM_{+CIM}^{+FC}}$
acquire a specific form in the case of consistified systems.

It should be clear from the construction of $\mathcal{R}^{\dagger}$
that the indexing sets $Q^{\dagger\left(\cdot,c\right)}$ of different
$\left(\cdot,c\right)$-bunches are disjoint, and that the union of
these indexing sets is the entire $Q^{\dagger}$ (consisting of all
$\xi=\left(q,c\right)$ such that $q\prec c$). This means that we
can use the same function to represent all $\left(\cdot,c\right)$-bunches,
\begin{equation}
R^{\dagger\left(\cdot,c\right)}=f\left(Q^{\dagger\left(\cdot,c\right)},X^{\left(\cdot,c\right)}\left(c\right)\right),
\end{equation}
where $X^{\left(\cdot,c\right)}\left(c\right)$ for different $c\in C$
is a set of stochastically unrelated random variables whose distributions
may vary with $c$. By forming an arbitrary coupling $X$ of $X^{\left(\cdot,c\right)}\left(c\right)$
for all $c\in C$, we can rewrite this as
\begin{equation}
R^{\dagger\left(\cdot,c\right)}=s\left(Q^{\dagger\left(\cdot,c\right)},\Proj_{\left(\cdot,c\right)}X^{\left(\cdot,c\right)}\right)=t\left(Q^{\dagger\left(\cdot,c\right)},X^{\left(\cdot,c\right)},\left(\cdot,c\right)\right),
\end{equation}
where $X^{\left(\cdot,c\right)}$ are stochastically unrelated distributional
copies of $X$. Since $Q^{\dagger\left(\cdot,c\right)}$ uniquely
determines $\left(\cdot,c\right)$, the function can be rewritten
as
\begin{equation}
R^{\dagger\left(\cdot,c\right)}=u\left(Q^{\dagger\left(\cdot,c\right)},X^{\left(\cdot,c\right)}\right).
\end{equation}
By the same argument, for all $\left(q,\cdot\right)$-bunches we have
\begin{equation}
R^{\dagger\left(q,\cdot\right)}=v\left(Q^{\dagger\left(q,\cdot\right)},Y^{\left(q,\cdot\right)}\right).
\end{equation}
The last two formulas represent the $\mathsf{HVM_{Gen}}$ for consistified
systems.

It can be easily shown that one can simplify this HVM by either making
the two functions $u$ and $v$ one and the same function or making
all $X^{\left(\cdot,c\right)}$ and $X^{\left(q,\cdot\right)}$ variables
identically distributed. For the latter option, create an arbitrary
coupling $\varPhi=\left(X,Y\right)$ and make its distributional copies
$\varPhi^{\left(\cdot,c\right)}$ and $\varPhi^{\left(q,\cdot\right)}$
for all contexts of $\mathcal{R}^{\dagger}$. Then
\begin{equation}
R^{\dagger\left(\cdot,c\right)}=u\left(Q^{\dagger\left(\cdot,c\right)},\Proj_{1}\varPhi^{\left(\cdot,c\right)}\right)=\phi_{1}\left(Q^{\dagger\left(\cdot,c\right)},\varPhi^{\left(\cdot,c\right)}\right),
\end{equation}
and
\begin{equation}
R^{\dagger\left(q,\cdot\right)}=v\left(Q^{\dagger\left(q,\cdot\right)},\Proj_{2}\varPhi^{\left(q,\cdot\right)}\right)=\phi_{2}\left(Q^{\dagger\left(q,\cdot\right)},\varPhi^{\left(q,\cdot\right)}\right).
\end{equation}
This is the form of the $\mathsf{HVM_{-CIM}^{+FC}}$ for consistified
systems: the distribution of the hidden variables is the same for
all contexts, but the observable variables depend on the type of the
context, $\left(\cdot,c\right)$-type or $\left(q,\cdot\right)$-type.
Thus, the $\mathsf{HVM_{-CIM}^{+FC}}$ representation for $R^{\dagger\left(\cdot,2\right)}=\left(R_{32}^{\cdot2},R_{42}^{\cdot2}\right)$
and $R^{\dagger\left(3,\cdot\right)}=\left(R_{31}^{3\cdot},R_{32}^{3\cdot},R_{33}^{3\cdot}\right)$
in system $\mathcal{R}_{0}^{\dagger}$ in (\ref{eq: exmaple consistified})
are, respectively:
\begin{equation}
\phi_{1}\left(Q^{\dagger\left(\cdot,2\right)},\varPhi^{\left(\cdot,2\right)}\right)=\left(\phi_{1}\left(\left(3,2\right),\varPhi^{\left(\cdot,2\right)}\right),\phi_{1}\left(\left(4,2\right),\varPhi^{\left(\cdot,2\right)}\right)\right)
\end{equation}
and
\begin{equation}
\phi_{2}\left(Q^{\dagger\left(3,\cdot\right)},\varPhi^{\left(3,\cdot\right)}\right)=\left(\phi_{2}\left(\left(3,1\right),\varPhi^{\left(3,\cdot\right)}\right),\phi_{2}\left(\left(3,2\right),\varPhi^{\left(3,\cdot\right)}\right),\phi_{2}\left(\left(3,3\right),\varPhi^{\left(3,\cdot\right)}\right)\right).
\end{equation}

The form of $\mathsf{HVM_{+CIM}^{-FC}}$ for consistified system obtains
by creating arbitrary couplings
\begin{equation}
\Psi_{1}=\left\{ R^{\dagger\left(\cdot,c\right)}:c\in C\right\} ,\Psi_{2}=\left\{ R^{\dagger\left(q,\cdot\right)}:c\in C\right\} ,
\end{equation}
and forming their distributional copies for all $\left(\cdot,c\right)$-bunches
and $\left(q,\cdot\right)$-bunches. Note that both $\Psi_{1}$ and
$\Psi_{2}$ are indexed by all $\left(q,c\right)\in Q^{\dagger}$.
Then
\begin{equation}
R^{\dagger\left(\cdot,c\right)}=\Proj_{Q^{\dagger\left(\cdot,c\right)}}\Psi_{1}^{\left(\cdot,c\right)}=\psi\left(Q^{\dagger\left(\cdot,c\right)},\Psi_{1}^{\left(\cdot,c\right)}\right)
\end{equation}
and
\begin{equation}
R^{\dagger\left(q,\cdot\right)}=\Proj_{Q^{\dagger\left(q,\cdot\right)}}\Psi_{2}^{\left(q,\cdot\right)}=\psi\left(Q^{\dagger\left(q,\cdot\right)},\Psi_{2}^{\left(q,\cdot\right)}\right).
\end{equation}
For our example with $R^{\dagger\left(\cdot,2\right)}=\left(R_{32}^{\cdot2},R_{42}^{\cdot2}\right)$
and $R^{\dagger\left(3,\cdot\right)}=\left(R_{31}^{3\cdot},R_{32}^{3\cdot},R_{33}^{3\cdot}\right)$
in (\ref{eq: exmaple consistified}), the $\mathsf{HVM_{+CIM}^{-FC}}$
representation is
\begin{equation}
\psi\left(Q^{\dagger\left(\cdot,2\right)},\Psi_{1}^{\left(\cdot,2\right)}\right)=\left(\psi\left(\left(3,2\right),\Psi_{1}^{\left(\cdot,2\right)}\right),\psi\left(\left(3,4\right),\Psi_{1}^{\left(\cdot,2\right)}\right)\right),
\end{equation}
\begin{equation}
\psi\left(Q^{\dagger\left(3,\cdot\right)},\Psi_{2}^{\left(3,\cdot\right)}\right)=\left(\psi\left(\left(3,1\right),\Psi_{2}^{\left(3,\cdot\right)}\right),\psi\left(\left(3,2\right),\Psi_{2}^{\left(3,\cdot\right)}\right),\psi\left(\left(3,3\right),\Psi_{2}^{\left(3,\cdot\right)}\right)\right).
\end{equation}

The falsifiable $\mathsf{HVM_{+CIM}^{+FC}}$, describing noncontextual
$\mathcal{R}^{\dagger}$ (hence also, noncontextual $\mathcal{R}$
in the CbD sense), is obtained by making $\mathsf{HVM_{+CIM}^{-FC}}$
and $\mathsf{HVM_{-CIM}^{+FC}}$ coincide:
\begin{equation}
R^{\dagger\left(\cdot,c\right)}=\psi\left(Q^{\dagger\left(\cdot,c\right)},\Psi^{\left(\cdot,c\right)}\right),
\end{equation}
\begin{equation}
R^{\dagger\left(q,\cdot\right)}=\psi\left(Q^{\dagger\left(q,\cdot\right)},\Psi^{\left(q,\cdot\right)}\right).
\end{equation}
Using again our example (\ref{eq: exmaple consistified}), the $\mathsf{HVM_{+CIM}^{+FC}}$
representation for $R^{\dagger\left(\cdot,2\right)}=\left(R_{32}^{\cdot2},R_{42}^{\cdot2}\right)$
and $R^{\dagger\left(3,\cdot\right)}=\left(R_{31}^{3\cdot},R_{32}^{3\cdot},R_{33}^{3\cdot}\right)$
is
\begin{equation}
\psi\left(Q^{\dagger\left(\cdot,2\right)},\Psi^{\left(\cdot,2\right)}\right)=\left(\psi\left(\left(3,2\right),\Psi^{\left(\cdot,2\right)}\right),\psi\left(\left(3,4\right),\Psi^{\left(\cdot,2\right)}\right)\right),
\end{equation}
\begin{equation}
\psi\left(Q^{\dagger\left(3,\cdot\right)},\Psi^{\left(3,\cdot\right)}\right)=\left(\psi\left(\left(3,1\right),\Psi^{\left(3,\cdot\right)}\right),\psi\left(\left(3,2\right),\Psi^{\left(3,\cdot\right)}\right),\psi\left(\left(3,3\right),\Psi^{\left(3,\cdot\right)}\right)\right).
\end{equation}

\section{Hidden assumptions about hidden variables?}

The literature on hidden variable models and contextuality contains
many attempts to explicate various assumptions underlying $\mathsf{HVM_{+CIM}^{+FC}}$.
We have seen that freedom of choice and context-independent mapping,
taken separately, are not assumptions, as they are universally satisfiable.
We have also seen that their conjunction is restrictive, but that
it is conceptually simpler to replace it with a single assumption,
one that I dubbed context-irrelevance. I will now briefly discuss
two additional propositions that are sometimes presented as assumptions.

\emph{Outcome determinism} is the assumption that hidden variables
and parameters of the situation (contents and contexts) uniquely determine
the observable outcomes. Some researchers find this assumption challengeable
\cite{Grnagier}. Did we not tacitly introduce this assumption somewhere
in the course of the development above? The answer is no: once one
consistently describes HVMs in the language of random variables, rather
than events and their probabilities, outcome determinism is satisfied
automatically. Unless one imposes constraints on the possible distributions
of $\Lambda^{c}\left(c\right)$, either of the two unfalsifiable HVMs
we have discussed, say, $\mathsf{HVM_{+CIM}^{-FC}}$, can be constructed
for any system of random variables. The very fact that the components
of $R^{c}$ are jointly distributed means that there is a random variable
of which all these components are measurable functions. This yields
the representation (\ref{eq:HVMCntInd}).

\emph{Factorizability} is another assumption that is often presented
as central for $\mathsf{HVM_{+CIM}^{+FC}}$ \cite{AbramskyBrand2011}.
Its meaning is that, using $\mathsf{HVM_{-CIM}^{+FC}}$ for definiteness,

\begin{equation}
p\left[\gamma\left(Q^{c},\Gamma^{c},c\right)=G\,|\,\Gamma^{c}=g\right]=\prod_{q\in Q^{c}}p\left[\Proj_{q\in Q^{c}}\gamma\left(Q^{c},\Gamma^{c},c\right)=\Proj_{q\in Q^{c}}\left(G\right)\,|\,\Gamma^{c}=g\right],
\end{equation}
where $G$ is a set of values indexed by $Q^{c}$, and $g$ is a specific
value of $\Gamma^{c}$. Did we not have to use this assumption? Within
our conceptual framework, we did not. Once outcome determinism is
accepted as trivially satisfied, factorizability has to be accepted
too. Indeed, all probabilities in this equation are equal 0 or 1,
and the left-hand side probability is 1 if and only if all the right-hand
side probabilities are 1.

\section{Conclusion}

Let us summarize.

\paragraph{\textmd{1. The propositions that are usually presented as the assumption
of free choice and the assumption of context-independent mapping in
constructing HVMs, when taken separately, are not in fact assumptions.
Rather they are two inter-translatable and universally satisfiable
ways of describing joint distributions of random variables in a system.
Because of their equivalence and their substantive emptiness these
notions are mere technical labels in HVMs: one should not take them
as saying anything about freedom of choice or physical influences
exerted by contexts in the sense in which these notions would be discussed
in science or philosophy.}}

\paragraph{\textmd{2. The conjunction of free choice and context-independent
mapping is a falsifiable (and de facto inapplicable to some systems)
model. However, rather than being a conjunction of two assumptions
(as they were viewed, e.g., in the historic discussion \cite{BellvsSHC1985}),
it is a single assumption in precisely the same sense in which a single
sentence can consist of two parts neither of which is a sentence.
One can avoid using the terminology of free choice and context-independent
mapping altogether, even as technical labels, by interpreting $\mathsf{HVM_{+CIM}^{+FC}}$
as an HVM with context-irrelevance: no distributions in this model
may depend on context.}}

\paragraph{\textmd{3. The positions just formulated are obtained almost automatically
if one systematically and carefully uses the language of random variables
in discussing HVMs. This also allows one to avoid the necessity of
certain additional assumptions, such as outcome determinism and factorizability.
To utilize the advantages of this language one has to pay meticulous
attention to the distinction between jointly distributed variables
and stochastically unrelated ones. ``Hidden variables'' are nothing
more than a tool for representing jointly distributed variables as
measurable functions defined on the same probability space --- which
is true essentially by definition. The variables from different contexts,
however, cannot be presented as functions of a single source of randomness,
even in the HVMs with context-irrelevance: the ``hidden variables''
in these models must still be indexed by contexts.}}

\paragraph{Acknowledgements}

This research was partially supported by the Foundational Questions
Institute grant FQXi-MGA-2201. I am grateful to Philippe Grangier,
Howard Wiseman, and Matthew Jones for thought-provoking discussions.
I am indebted to Jerald D. Balakrishnan for critically reading an
earlier draft of this paper and suggesting many improvements. I am
grateful to Samson Abramsky, Ad\'an Cabello, and Pawe\l{} Kurzy\`nski
for co-organizing with me the meeting on contextuality (QCQMB 2022)
at which some of the issues presented in this paper have been the
subject of lively discussions.

\end{document}